\documentclass{aipproc}
\layoutstyle{6x9}

\newcommand{\be}{\begin{eqnarray}}
\newcommand{\ee}{\end{eqnarray}}
\newcommand{\bc}{\begin{center}}
\newcommand{\ec}{\end{center}}

\begin{document}

\title{Fishing in black holes}


\author{A. Brotas \footnote{brotas@fisica.ist.utl.pt}}{
  address = {Departamento de F\'{\i}sica, Instituto Superior T\'ecnico, \\
             Av Rovisco Pais, 1096 Lisboa Codex, Portugal}}


\begin{abstract}

The  coordinate system $ \left( \bar{x} , \bar{t} \right) $  defined by: $r = 2m + K \bar{x} - c K ~ \bar{t} ~~~~$
 and \\
 $ t = \frac{\bar{x} }{c K } - \frac{1}{c K} \int_{r_a}^{r}\left(1-
\frac{2m}{r} + K^2 \right)^\frac{1}{2}  \left( 1 - \frac{2m}{r} \right)^{-1} dr $
allow us to write  the Schwarzschild metric \\ 
in the form:
$ ds^2 = c^2 d\bar{t}^2 + \left( \frac {W^2}{K^2} - \frac{2W}{K} \right)
 d\bar{x}^2 + 2 c \left( 1 + \frac{W}{K} \right) d\bar{x} d\bar{t} -
r^2 \left( d\theta^2 + \cos^2\theta \:  d\varphi^2 \right) $ \\
with: $  W = \left( 1 - \frac{2m}{r}  + K^2 \right)^\frac{1}{2} $
in which the coefficient's pathologies are moved
to $ r_K = \frac{2m}{1+ K^2} $. 

This new coordinate system  is used to study the entrance in a black hole
 of a rigid line (a line in which the shock waves propagate with velocity c ).

\end{abstract}

\maketitle
\section{The fishing problem}

A fisherman caught a fish with a hook on a line but he let it cross a Schwarzschild horizon.
We know that the fish can never leave the black hole again.
Therefore the fisherman will not be able to retrieve it.
What will happen when he pulls the line?
Admitting that the line is undeformable, it would seem that the
fish should pull the fisherman into the black hole.
Is that so?

Can the fisherman remain outside the black hole ?
Or will he be inexorably dragged by the fish if he remains attached to it through the line?

To clarify this issue we need to revue some notions:

1- In relativity it is necessary to make a distinction between "rigid bodies"
and "undeformable bodies: "rigid body" is a physical concept and "undeformable body" is a geometrical one.
In a textbook of Physics the word "rigid" means
 always: "as rigid as possible" (rigidity limit).

2- In classical Physics books rigid bodies are undeformable
but in relativity a "rigid body" must be a body in which shock waves propagate with maximun speed c . Such a body is deformable .

3- In 1909, Born studied the undeformable relativistic body but he
 made the mistake of calling it rigid \cite{um}.
The "rigid body" that we find today in the books on Relativity is
 the undeformable body described by Born in 1909.
This error of designation was the cause of numerous paradoxes
and set back the discovery of the elastic laws of rigid bodies.

These very curious and practically unknown laws, discovered, in 1952, by Mc Crea and Hogart
(in the one dimension case) \cite{dois} and later, in 1968, by Ant\'onio Brotas \cite{tres}
\cite{quatro}, are the following:
$p= \frac{\rho^0_0 c^2}{2} \left(\frac{1}{s^2} - 1\right) ~~~~~
\rho_0 = \frac{\rho^0_0}{2} \left(\frac{1}{s^2} + 1\right) $. \\

These laws enlighten us regarding the fisherman's problem. The line that links the fisherman to the fish is 
a physical object. At most, it could be a rigid line that might be stretched infinitely.
As long as the fisherman has enough force to hold the line
 $ \left( f> \frac{\rho_o^o c^2}{2} S \right) $      
  he will not be pulled into the black hole.
\section {The differential equations}
In relativity , in Minkowski space , the elasticity laws of rigid bodies and the conservation laws give us the equation:
\be \frac{\partial^2 X}{\partial x^2}-\frac{1}{c^2} \frac{\partial^2 X}
{\partial t^2} = 0 ~~~ .\ee
This equation has the solutions: $  X = g \left(x - ct\right) + h \left(x+ c t \right) =
 g \left(o_1\right) + h \left( o_2\right) $. 

In the Schwarzschild metric the equation of the radial motion of a rigid bar with the usual coordinates $(r,t)$
 takes the form:
\be \frac{\partial^2 X}{\partial r^2}-\frac{\frac{\partial^2 X}
{\partial t^2}}{c^2 \left(1-\frac{2m}{r} \right)^2} + 
\frac{\frac {2m}{r^2} \frac{\partial X}{\partial r}}{\left( 1-\frac{2m}{r} \right)} = 0 ~~~ , \ee
which has solutions of the type:
 \be X = g \left( r + 2m \log \left| r-2m \right| - ct \right)  +
 h \left( r + 2m \log \left| r-2m \right| +ct \right) =
	g\left( O_1 \right) + h \left( O_2 \right) ~~ . \ee

The point $ r=2m $ is a point of no return where the arguments of the functions $g$ and $h$ have
infinite values.
 However this difficulty can be removed by choosing another coordinate system.
\section {A coodinate system associated with an undeformable line}
We cannot use an undeformable line to angle a fish but we may use it to define a 
coordinate system. We will use a new coodinate system $ \left( \bar{x},\bar{t} \right) $ 
where $ \bar{x} $ is the length of an undeformable line pulled by a winch located at a point $r_a$ and $\bar{t}$
is the proper time displayed by clocks conveniently synchronized and fixed to the point's line.
These coordinates are related to usual coordinates $\left( r,t \right) $ by the formula:
\be \begin{array}{ccc} r = 2m + K \bar{x} - c K \bar{t} & ; &
 t = \frac{\bar{x} }{c K } -
\frac{1}{c K} \int_{r_a}^{r}{\left(1- \frac{2m}{r} + K^2 \right)^\frac{1}{2}
 \left( 1 - \frac{2m}{r} \right)^{-1} dr} \end{array} ~~ . \ee
   The length $ \bar{x} $  of the line which comes out of the winch within the interval  $ dt $  is   $ d\bar{x} = c K dt$.
 ($ K $ is a constant associated with the velocity of the winch: $ 0 < K < \infty $ ).
 
With this new coordinate system the Schwarzschild metric takes the form:
\be  ds^2 = c^2 d\bar{t}^2 + \left( \frac {W^2}{K^2} - \frac{2W}{K} \right)
 d\bar{x}^2 + 2 c \left( 1 + \frac{W}{K} \right) d\bar{x} d\bar{t} -
r^2 \left( d\theta^2 + \cos^2\theta  d\varphi^2 \right) ~~ , \ee
 with: $~~ W = \left( 1 - \frac{2m}{r}  + K^2 \right)^\frac{1}{2} $, 
 in which the coefficient's pathologies are moved to $r_K = \frac{2m}{1+K^2}$.
 (The derivation/justification of these formula is somewhat complicated , but their verification/confirmation
is a mere mathematical routine matter).

Direct computation of the Ricci tensor for the metric (5) gives
 $ R_{\alpha \beta} = 0 $ as it should be for the Schwarzschild metric.

\section {The equation and the solutions in this new system}
In this new coordinate system, the prior Eq.(2) of radial motion of the rigid line takes the form:
\be \frac{\partial^2 X}{\partial \bar{x}^2}  + 
\frac{1}{c^2} A  
\frac{\partial^2 X}{\partial \bar{t}^2} - 
\frac{2B}{c^2} \frac{\partial^2 X}{\partial\bar{x} \partial\bar{t}} + \frac{ m }{r^2W} 
\frac{\partial X}{\partial\bar{x}} +\frac{m}{c r^2} \left[ \frac{1}{W} -\frac{2}{K} \right]
\frac{\partial X}{\partial \bar{t}} =0 ~~ , \ee
with $ A= g_{\bar{x}\bar{x}} = \frac{W^2}{K^2}- \frac{2W} {K} ~~~ and ~~~
  B = g_{\bar{x}\bar{t}} = -c + c \frac{W}{K} $.
This equation is rather complicated, but it has the advantage of its coefficients having 
no singularities for $ r > \frac{2m}{1+K^2} $. \\

And how about the solutions? How are represented in the new coordinates the solutions of the type :
 $ X =h \left( r + \log \left| r-2m \right| +ct \right) =
  h \left( O_2 \right) $ ?

We may write: $~~ O_2 = 2m + \left( K + \frac{1}{K} \right) \bar{x} - c K \bar{t}+
2mY - \frac{I}{K} $ , \\
with: $ 2mY = 2m \log \left| r-2m \right|  =
2m \int_b ^W \left(\frac{1}{W-K} + \frac{1}{W+K}  + \frac{-1}{W-n} + \frac{-1}{W+n} \right) dW $ , \\
and $ \frac{I}{K}  = \frac{m}{K} \int_b ^W \left(\frac{2K}{W-K} - \frac{2K}{W+K}
+\frac{e}{W+n}  + \frac{-e}{W-n} + \frac{1 }{ \left(W+n\right)^2} + \frac{1 }{ \left(W-n\right)^2}\right)dW $  , \\
( with: $ a= 1 + 2m $ ; $ b = \left(\frac{1}{a} + K^2 \right)^\frac{1}{2}$
 ;  $ n = \left( 1 + K^2 \right)^\frac{1}{2}$ ; $ e = \frac{1+2K}{n}$) . \\

The divergent terms  in $ 2mY $    and $ \frac{I}{K}$ for $W = K$ $(i.e ~~ r=2m)$ cancel for $O_2$.

We checked directly (with many days of work) that  $ X = H \left ( \bar{x}, \bar{t} \right) =
 h \left( O_2 \left( \bar{x}, \bar{t} \right) \right) $
effectively is a solution of Eq.6 for $ r > \frac{2m}{1+K^2} $ .

And how about the solutions: $ X = g \left( r + 2m \log \left| r-2m \right| -  ct \right) =
  g \left( O_1 \right) $ which correspond to lines moving away from the black hole?

With similar computations we obtain: $ O_1 = 2m + \left( K - \frac{1}{K} \right) \bar{x} -
c K \bar{t} + 2mY + \frac{I}{K} = r - \frac{\bar{x}}{K} +
 m \int_b ^W \frac{4 dW}{W-K} + R ~~~~$ with:
$ R = 4m + \int_{-b} ^W \left( \frac{-W-b}{W^2 - n^2} +
\frac{n^2}{K \left(W^2 -n^2\right)^2}	\right) dW  $

Now the problem is more complicated because for $W = K$ $(i.e ~~ r=2m)$
 the argument  of   g  is $ - \infty $ .
This means that we cannot get a solution to Eq.(5) through a mere substitution of variables in the
solutions  $ g \left( O_1 \right) $. However, considering the "function":
\be f^{\ast} \left(O_1 \right) =  \pm \exp \left( \frac{O_1}{4m}  \right) ~~~~~;~~
\left( + ~~\ if ~~\ W>K ~~ and ~~ - ~~\ if ~~\ W>K \right) ~~ , \ee
we can write:
$ X = g^{\ast} \left( \bar{x}, \bar{t} \right) =
  f^{\ast} \left(O_1 \right) =
 \frac{W-K}{b-K} \exp \left( - \frac{ \bar{x}}{4mK} + 
  \frac{r+R}{4m} \right) $. \\

This function $ X = g^{\ast} \left( \bar{x}, \bar{t} \right)$, which is properly defined
and differentiable  for $ r > r_K$, is a solution of Eq.(6).
We checked this directly.
The same happens with: 
\bc $ X = G^{\ast} \left( \bar{x}, \bar{t} \right) = \left[ a +
 j \left( g^{\ast} \left( \bar{x}, \bar{t} \right) \right) \right] 
g^{\ast} \left( \bar{x}, \bar{t} \right)$ ~~ , \ec
provided that the conditions $ j \left(0 \right) = 0 $ and 
 $ j' \left(0 \right) = 0 $ are fulfilled. \\ 
 These solutions correspond to lines with a
 fixed point at the horizon $ r_h = 2m $  which are stretched on both sides.

Summarizing , we may say that the functions:
$ X = G^{\ast} \left( \bar{x}, \bar{t} \right) +
 H^{\ast} \left( \bar{x}, \bar{t} \right) $ are solutions of Eq.(6).
\section {Crossing the horizon}
Taking a specific solution, we can evaluate the velocity and the deformation 
(and consequently stress) of the rigid line at different points and moments in its
evolution, provided $r > r_K = \frac{2m}{ 1 + K^2}$.

We verify that when crossing the
Schwarzschild horizon the points velocity is within the extreme cases of the fixed point of the solutions $G^*$
and the velocity $-c$ from solutions $H^*$. Concerning the deformation and the stress, they may be either zero,
a compression, or an expansion (traction).

Based on these results we may question the generally accepted statement that 
an extended body disintegrates when entering a black hole.

Our idea is that if it was possible to experience with our body the internal 
stress of a rigid bar entering a black hole we would feel nothing crossing the Schwarzschild
horizon.



\end{document}